# Mode-locked and tunable fiber laser at the 3.5 μm band using frequency-shifted feedback

Ori Henderson-Sapir,[1,2,†*] Nathaniel Bawden,[1,†] Matthew R. Majewski,[3] Robert I. Woodward,[3] David J. Ottaway,[1] Stuart D. Jackson[3]

[1] Department of Physics and Institute of Photonics and Advanced Sensing, The University of Adelaide, Adelaide, Australia
[2] Mirage Photonics, Oaklands Park, Australia
[3] MQ Photonics, School of Engineering, Faculty of Science and Engineering, Macquarie University, North Ryde, Australia
[†]These authors contributed equally to this Letter.
*Corresponding author: ori.henderson-sapir@adelaide.edu.au

**We report on a mid-infrared mode-locked fiber laser that uses an acousto-optic tunable filter to achieve frequency-shifted feedback pulse generation with frequency tuning over a 215 nm range. The laser operates on the 3.5 μm transition in erbium-doped zirconium fluoride-based fiber and utilizes the dual-wavelength pumping scheme. Stable, self-starting mode-locking with a minimum pulse duration of 53 ps was measured using a two-photon absorption autocorrelator. The longest wavelength demonstrated was 3612 nm and a maximum average power of 208 mW was achieved. This is the longest wavelength rare-earth doped mode-locked fiber laser demonstrated to the best of the authors' knowledge. The broad tunability promises potential uses for environmental sensing applications.**

In recent years, the performance demonstrated by mid-infrared fiber lasers has increased significantly and the wavelength achieved from room temperature devices has been pushed deeper into the infrared. [1]. This trend towards longer wavelengths has also been occurring for rare-earth-doped, mode-locked fiber lasers. A first demonstrations by Frerichs at 2.8 μm using erbium-doped fiber was in 1996 with Q-switched mode-locking using saturable absorbers [2]. In 2012 Li demonstrated CW mode-locking using a holmium/praseodymium co-doped fiber and a semiconductor saturable absorber mirror [3]. Further reduction in the pulse duration from picosecond to femtosecond time scale required utilizing virtual saturable absorbers in the form of non-linear polarization evolution (NPE) techniques to achieve shorter pulses, of 497 fs and later 230 fs, respectively [4, 5] for wavelengths around 2.9-3.0 μm using erbium-doped mode-locked fiber lasers. Further work increased the wavelength of rare-earth-doped, mode-locked fiber lasers to the 2.97-3.30 μm range using dysprosium doped fibers [6-8].

Continued extension of the wavelength coverage is enabled by the $^4F_{9/2} \rightarrow {}^4I_{9/2}$ transition in erbium emitting in the 3.5 μm band. A mode-locked laser operating around 3.5 μm would have many potential applications in spectroscopy and environmental sensing [9] due to strong optical activity of many greenhouse gasses and a number of volatile organic compounds. Wavelength tunabilty would further increase the utility and selectivity of such methods.

Thus far the operation of these sources has relied on novel saturable absorber materials. The first Q-switched, mode-locked erbium fiber laser in this spectral region used black phosphorous [10]. However, the mode-locked pulse duration was not explicitly measured for this work so little can be said about its pulse parameters.

Another promising material is graphene, and there has been significant progress in utilizing its broadband saturable absorber properties for achieving ultrafast CW operating mode-locked lasers in the near infrared. However, this success has not been reproduced in the mid-infrared beyond 2.8 μm. The only demonstration of a mid-infrared mode-locked laser using graphene achieved pulses of 65 ps at the 2.8 μm band [11]. Our own attempts to obtain short-pulse mode-locking using graphene at 3.5 μm as a saturable absorber in an erbium-doped zirconium fluoride based ($ZrF_4$) fiber laser were not successful. Further investigations suggest strong two photon absorption at wavelengths beyond 3 μm significantly reduced the achievable modulation depth, thereby preventing short pulse mode-locking [12].

Recently, frequency-shifted feedback (FSF) techniques have been shown to be a favorable alternative to saturable absorbers and have been used to demonstrate ps-scale pulsed lasers at the high energy portion of the mid-infrared around 3 μm [6-8]. Pulses with 33 ps duration were recently demonstrated in a dysprosium-doped $ZrF_4$ fiber laser that used an acousto-optical tunable filter (AOTF).

Tuning over 2.97 to 3.3 µm was achieved [8]. An order of magnitude shorter pulses of 4.7 ps were obtained in a holmium/praseodymium co-doped ZrF$_4$ fiber using an acousto-optical modulator (AOM) [7].

FSF is a technique where intracavity frequency shifting in conjunction with filtering is introduced [13]. At every round trip the light frequency red-shifts by twice the drive frequency of the AOTF or AOM, eventually shifting after a number of round trips beyond the bandwidth of the filter thus suppressing coherent continuous-wave operation. Pulsed operation can result, however, if there is sufficient cavity non-linearity. In this case intense pulses undergo self-phase modulation that broaden the spectrum, pushing a portion of the pulse energy back into the filter passband, counteracting the AOTF induced frequency shift. The result is higher net gain for a pulsed state leading to a pulse train with a frequency equal to that of the cavity free spectral range (FSR). We note that while this is not 'mode-locking' in the traditional sense whereby the longitudinal modes are locked in phase, the literature however has adopted the term to reflect the strong similarities in the output characteristics, and we follow this convention here.

In this paper we extend the wavelength range achieved using the FSF technique further into the mid-infrared to the 3.5 µm band. Using an erbium-doped ZrF$_4$ fiber and utilizing the dual-wavelength pumping technique (DWP) we demonstrate over 200 nm of wavelength tunability at this band while simultaneously demonstrating mode-locked operation up to 3612 nm, the longest wavelength from a rare-earth-doped, mode-locked fiber laser demonstrated to-date.

The now well established DWP technique, shown schematically in the energy level diagram in Fig. 1, uses two pump wavelengths to facilitate efficient lasing on the $^4F_{9/2} \rightarrow {}^4I_{9/2}$ transition in erbium [14, 15]. Light from a 977 nm laser diode first excites erbium ions from the ground state to the $^4I_{11/2}$ long-lived state. The ions are then excited a second time to the upper lasing level $^4F_{9/2}$ using light from a 1973 nm thulium-doped fiber laser. Ions that contribute to lasing return to the $^4I_{11/2}$ level, where they can be re-pumped by the 1973 nm pump light.

The experimental setup for our laser system is shown schematically in Fig. 1. P$_1$ is a multimode commercially available 977 nm fiber-coupled laser diode (LIMO HLU30F200-977) and P$_2$ was an in-house built thulium-doped silica fiber operating at 1973 nm, as in Ref. [15]. The light from both pumps was combined using free-space optics (DM1 in Fig.1) and mode matched into the fiber cladding and core, respectively.

The pumps were passed through a 45° dichroic mirror (DM2) which was highly transmissive (HT) for both pump wavelengths and highly reflective (HR) at the 3.5 µm band. Both pumps were then launched into the ZrF$_4$ fiber through a custom made ZnSe asphere. The gain medium was a 2.8 m double clad 1 mol.% erbium-doped ZrF$_4$ glass fiber. The fiber (Le Verre Fluore, France) had a 16 µm core, a 240/260 double truncated circular inner cladding with a measured NA of 0.08 at 3.5 µm. The fiber distal end was butted against a dichroic mirror (DM3) which is 95% reflective at 3.5 µm. The output beam from this side was used mostly for diagnostics and to aid during the alignment process and did not affect the laser power significantly.

On the pump input end of the fiber the aspheric lens collimated the 3.5 µm intracavity mode and the beam was reflected by DM2 mirror onto the AOTF. The AOTF (Gooch & Housego) consisted of a TeO$_2$ crystal in a quasi-collinear configuration with a tunable center frequency determined by the radio frequency (RF) driver frequency. The resonator was completed by a silver mirror providing the feedback for the AOTF first diffracted order, while the zero order was used for output coupling the laser. The laser output through DM3 was used for diagnostic purposes with the beam going to a thermal power meter, monochromator, thermal camera or an autocorrelator. A 3 µm longpass filter was used on both outputs to filter residual pump light.

In order to optimize the intracavity operation, it was necessary to characterize the AOTF, the relationship between the drive frequency and central wavelength of the filter as well as the optimal driver power for each wavelength, see Fig. 2. A Mirage Photonics MFL-3500 tunable mid-infrared fiber laser was used for this characterization [16]. The beam from the MFL-3500 was collimated and split into two arms: the first directed to a reference detector and the other through the AOTF aperture. The laser wavelength was varied and at each wavelength the AOTF drive frequency was changed until maximum transmission into the first diffracted order was obtained. The bandwidth of the AOTF was found by measuring the relative transmission through the AOTF when the input laser wavelength was 3471 nm, the RF power fixed and the drive frequency was varied, see Fig. 2. inset.

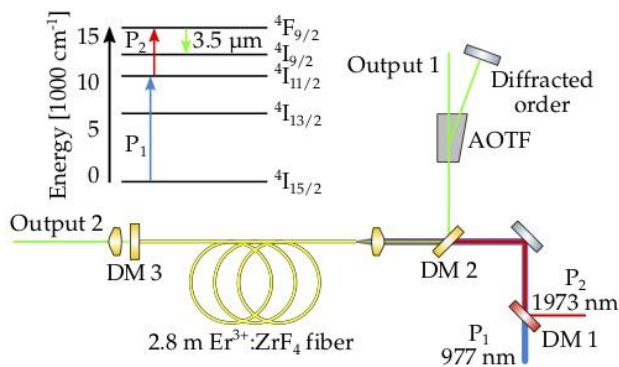

Fig.1. *Experimental setup of the dual-wavelength pumped frequency-shifted feedback tunable fiber laser. DM – dichroic mirror. Inset: Simplified energy level diagram.*

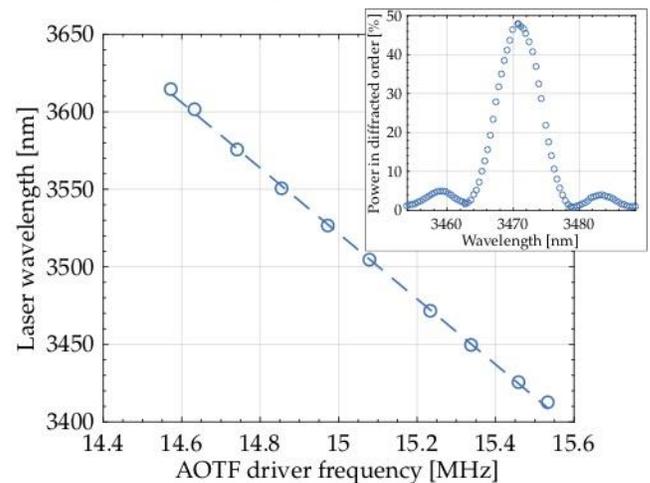

Fig. 2. *Measured AOTF characteristics: Wavelength of transmission as a function of drive frequency. Inset: AOTF transmission curve at one RF drive frequency operating at 260 mVpp.*

Lasing threshold for 3471nm wavelength was reached with $P_1$ and $P_2$ levels of 3.2 W and 3.0 W, respectively. Subsequent measurements used a fixed $P_1$ of 4.8 W while varying $P_2$ up to a maximum incident power of 4.9 W. A maximum average power of 208 mW was achieved when both outputs were summed, as shown in Fig. 3. The AOTF zero order output had four times higher power than the output through the 95% reflective DM3 dichroic. The AOTF diffraction efficiency was saturated at an estimated 70% based on prior extra-cavity measurement and well below its maximum drive voltage. A slope efficiency of 8.4% total from both outputs was measured relative to incident $P_2$ with $P_1$ power fixed at 4.8 W.

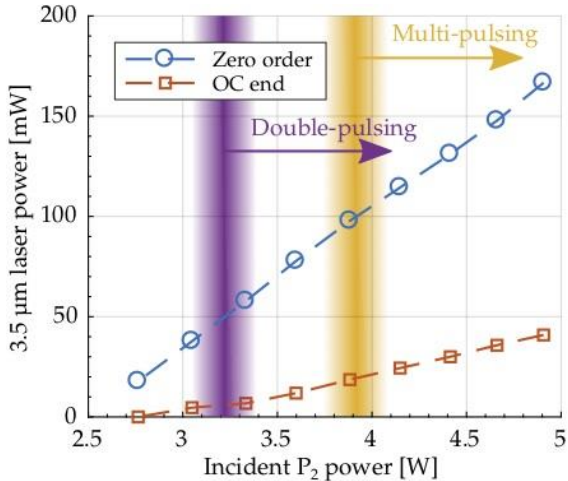

Fig. 3. Measured laser average output power. Slope efficiency of 8.4% is obtained relative to $P_2$ with a fixed $P_1$ of 4.8 W.

Q-switched mode-locking [8, 13] or CW-like behavior [7] is often observed in AOTF based FSF systems upon reaching threshold. However, we did not observe either of these phenomena and stable, single-pulse mode-locking was observed once the laser achieved threshold. At higher pump powers, multi-pulsing behavior was observed using a fast detector (VIGO PEMI) and an oscilloscope. Two, three and even four pulse trains can be seen propagating simultaneously within the resonator, see Fig. 4. The temporal separation between the subsequent pulses suggest that they are due to reflection off the fiber flat cleave at the pump end. Angle cleaving was attempted but introduced high losses that prevented the system from reaching threshold.

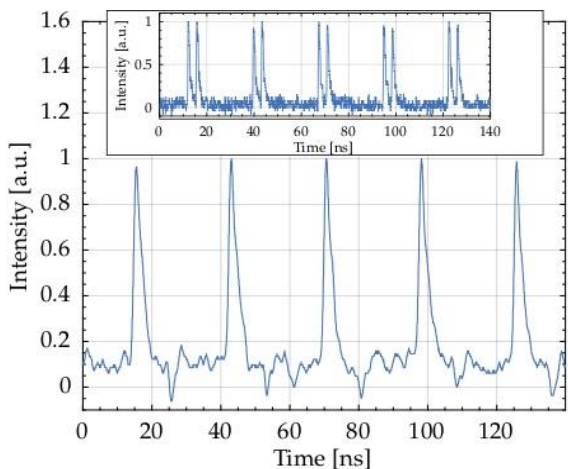

Fig. 4. Measured mode-locked pulse train. Inset: Double pulsing.

In the single pulse regime, the repetition rate was 36.23 MHz in agreement with the inverse round-trip time of the resonator. In the RF spectrum domain, seen in Fig. 5., a clear pulse train extending to the detector bandwidth limit at ~1 GHz was present. When inspecting using a narrower frequency span, a clean, fundamental beat note at the inverse cavity round trip time is observed. The peak is 53 dB above the pedestal when using a 100 Hz resolution bandwidth. Side lobes to the relaxation oscillation frequency were observed, 80 kHz offset from the fundamental. The energy per pulse while still in the single pulse regime was 1.38 nJ.

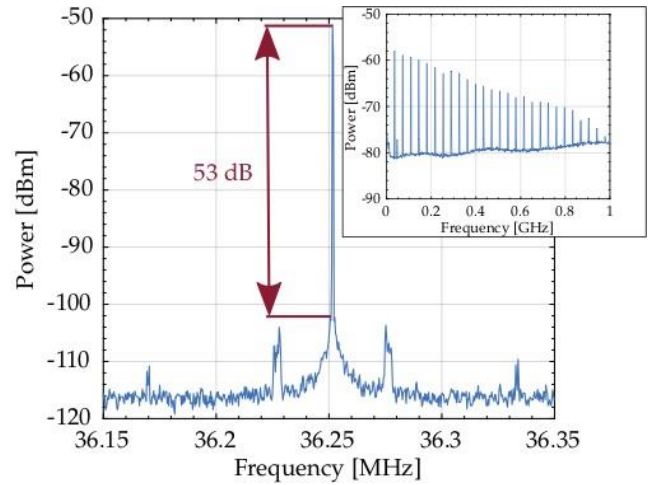

Fig. 5. Measured RF spectrum of the laser output. The side lobes at 80 kHz from the fundamental mode are relaxation oscillations mixing with the signal carrier. Inset: Wideband. The small spur next to the fundamental is due to electronics noise.

The optical spectrum exhibited the typical shoulder observed in other FSF based mode-locked fiber lasers [8, 17]. A full width at half maximum (FWHM) linewidth of $0.58 \pm 0.03$ nm was observed at all power levels and wavelengths of operation, see Fig. 6. This linewidth is twice the observed linewidth when using a diffraction grating [15]. This spectral bandwidth implies a pulse width of 30 ps, assuming a transform-limited Gaussian pulse. Most commercial autocorrelators cannot measure such a long pulse. We therefore built an autocorrelator based on two-photon absorption, similar to [8] in a Michelson interferometer to characterize the pulse length. An example of an autocorrelator trace is shown in Fig. 7., where pulses of 53 ps were measured. This pulse duration significantly exceeds the transform-limited duration under the assumption of a Gaussian pulse shape, indicating strong pulse chirp. However, prior studies have shown that FSF lasers often produce asymmetric non-Gaussian pulses [8, 17] (which cannot be fully resolved by an autocorrelator), thus it is an area of further work to investigate the exact pulse shape and to directly quantify the chirp.

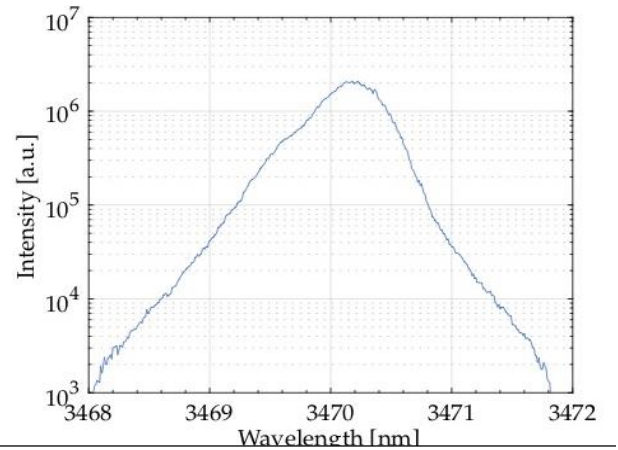

Fig. 6. Measured laser spectrum of the mode-locked system. The typical "shoulder" of FSF lasers' spectrum is evident. The FWHM at all power levels was measured to be $0.58 \pm 0.03$ nm.

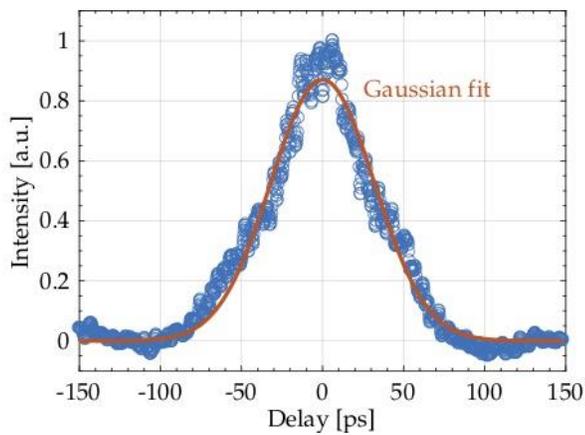

Fig. 7. Measured autocorrelation trace using an intensity autocorrelator employing two-photon absorption. The deconvolved pulse width is 53 ps assuming a Gaussian pulse.

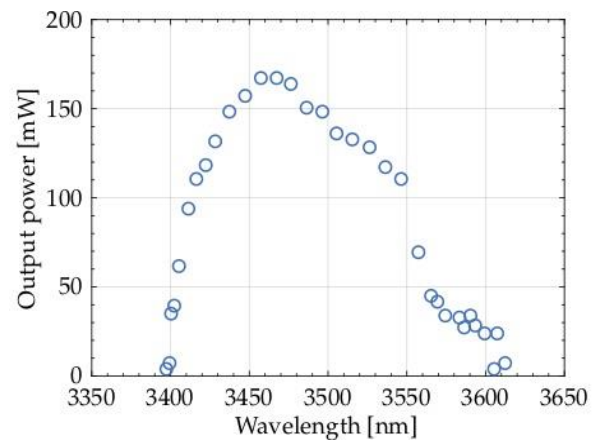

Fig. 8. Measured emission wavelength coverage of the laser. Average power measured at the AOTF zero diffraction order output. The diffraction efficiency was maximized with the AOTF operating close to its maximum RF driver current. 20% additional power emerged out of the DM3 dichroic end.

AOTF-based, FSF mode-locked systems can have pulse durations that are long compared to traditional mode-locking techniques because of the bandwidth broadening limitations imposed by the narrow bandwidth of the quasi-collinear configuration of the AOTF, see Fig. 2. The 53 ps long pulses are therefore reasonable when compared to the FSF based dysprosium fiber laser demonstrated by Woodward *et al.* [8]. Furthermore, the gain of the $^4F_{9/2} \rightarrow ^4I_{9/2}$ transition in erbium is significantly lower than the dysprosium laser which means that only wavelengths very close to the center of the filter have enough gain to oscillate.

Maximum wavelength tuning of 215 nm was achieved at maximum pump power levels of 4.8 W and 4.9 W for $P_1$ and $P_2$, respectively, see Fig. 8. The AOTF RF driver was pushed to the maximum power permissible for the AOTF. The bandwidth achieved is less than half that achieved with direct tuning using a grating in Littrow configuration [15]. The reasons for the reduced bandwidth are still unclear, however there are multiple likely possibilities including increased losses due to the strong polarizing effect of the AOTF, the AOTF loss of 37% for a double pass for linearly polarized light, and the fact that the current laser configuration (pump and the laser output coming from the same side) requires precise mode matching to ensure optimal pump launch while maintaining perfect collimation of the 3.5 μm light onto the AOTF. Since a single ZnSe asphere was used it was not possible to simultaneously satisfy precisely all three wavelengths i.e., $P_1$, $P_2$ and 3.5 μm Therefore, the asphere's position was optimized initially for $P_2$ launching and after obtaining lasing its position was optimized for maximum power at 3.5 μm.

In conclusion, we have demonstrated the longest operating wavelength for a mode-locked rare-earth doped fiber laser. The laser employed the frequency-shifted feedback method on the 3.5 μm transition in erbium-doped $ZrF_4$ fiber while using dual wavelength pumping. A maximum average output power of 208 mW was demonstrated with 53 ps long pulses at a repetition rate of 36.23 MHz. In addition, the laser exhibited 215 nm of wavelength tuning. The frequency-shifted feedback technique is demonstrated to be an effective method for obtaining electronic wavelength tunability in the mid-IR part of the spectrum where many important molecules have strong optical activity.

**Funding.** South Australian Government Premier's Research and Industry Fund (PRIF grant); Asian Office of Aerospace R&D (AOARD) Grant FA2386-19-1-0043

**Acknowledgment**. The authors acknowledge the expertise, equipment, and support provided by the Australian National Fabrication Facility (ANFF) at The University of Adelaide. This research was supported by the South Australian Government Premier's Research and Industry Fund (PRIF), Australian Research Council and the Asian Office of Aerospace R&D.

**Disclosures**. OHS: Mirage Photonics (I,P)